\documentstyle[preprint,prl,aps,epsf]{revtex}
\begin{document}
\newcommand\ie {{\it i.e. }}
\newcommand\eg {{\it e.g. }}
\newcommand\etc{{\it etc. }}
\newcommand\cf {{\it cf.  }}
\newcommand\etal {{\it et al. }}
\newcommand{\be}{\begin{eqnarray}}
\newcommand{\ee}{\end{eqnarray}}
\tightenlines
\def\lsim{\mathrel{\raise.3ex\hbox{$<$\kern-.75em\lower1ex\hbox{$\sim$}}}}
\def\gsim{\mathrel{\raise.3ex\hbox{$>$\kern-.75em\lower1ex\hbox{$\sim$}}}}
\vskip 2.5cm

\title{Inhomogeneous Shadowing Effects on $J/\psi$ Production in $dA$
Collisions}

\author{S. R. Klein$^1$ and R. Vogt$^{1,2}$}

\address{{
$^1$Nuclear Science Division, Lawrence Berkeley National Laboratory, 
Berkeley, CA
94720, USA\break 
$^2$Physics Department, University of California, Davis, CA 95616, USA}\break}

\vskip .25 in
\maketitle
\begin{abstract}

We study the effect of spatially homogeneous and inhomogeneous
shadowing on $J/\psi$ production in deuterium-nucleus collisions.  We
discuss how the shadowing and its spatial dependence may be measured
by comparing central and peripheral $dA$ collisions.  These event
classes may be selected by using gray protons from heavy ion breakup
and events where the proton or neutron in the deuterium does not
interact.  We find that inhomogenous shadowing has a significant
effect on central $dA$ collisions, larger than is observed in central
$AA$ collisions.  The inhomogeneity may be measured by comparing the
rapidity dependence of $J/\psi$ production in central and peripheral
collisions.  Results are presented for $d$Au collisions at
$\sqrt{s_{NN}} = 200$ GeV and $d$Pb collisions at $\sqrt{s_{NN}} =
6.2$ TeV.

\end{abstract}

\section{Introduction}

The nuclear quark and antiquark distributions have been probed through
deep inelastic scattering (DIS) of leptons and neutrinos from nuclei.
These experiments showed that parton densities in free protons are
modified when bound in the nucleus \cite{Arn}.  This modification,
referred to collectively as shadowing, depends on the parton momentum
fraction $x$ and the square of the momentum transfer, $Q^2$.  Most
models of shadowing predict that the modification should vary
depending on position with in the nucleus\cite{ekkv4}.  Most DIS
experiments have been insensitive to this position dependence.
However, some spatial inhomogeneity has been observed in $\nu N$
scattering in emulsion \cite{E745}.

Deuterium-nucleus collisions at heavy ion colliders offer a way to
measure the structure functions of heavy nuclei at higher energies,
and hence lower $x$ and higher $Q^2$ than are currently possible with
fixed target DIS experiments.  In addition, $dA$ collisions are more
sensitive to the gluon distributions in nuclei than DIS.  
These $dA$ collisions are preferred over $pA$ collisions for
technical reasons - the two beams have similar charge $Z$ to mass $A$
ratios, simplifying the magnetic optics.  Several groups have
previously considered structure function measurements using $pA$ and
$dA$ collisions.  However, the spatial dependence has not yet been
explored.  A measurement of the spatial dependence would be of great
interest both as a probe of the shadowing mechanism and as an
important input to studies of $AA$ collisions.

In this letter, we show that $dA$ collisions can be used to study the
spatial dependence of nuclear gluon shadowing.  We consider two
concrete examples: $d$Au interactions at $\sqrt{s_{NN}} = 200$ GeV at
RHIC and 6.2 TeV $d$Pb interactions at the LHC.  The RHIC conditions
correspond to the Spring, 2003 run.

We choose the $J/\psi$ as a specific example since it has already been
observed in $pp$ and Au+Au interactions at RHIC \cite{frawley}.  Since we wish
to focus on the magnitude of the shadowing effect, we do
not include nuclear absorption or transverse momentum, $p_T$, broadening
in our calculations.  The techniques and general conclusions that we
discuss should be equally applicable to other hard probes such as $\Upsilon$,
open charm and bottom quarks, Drell-Yan dileptons, jets, and dijets.
At the LHC, similar approaches should also apply for $W^{\pm}$ and
$Z^0$ production.  By using a variety of probes, both quark and gluon
shadowing can be measured over a wide range of $x$ and $Q^2$.

Our $J/\psi$ calculations employ the color evaporation model (CEM) which 
treats all
charmonium production identically to $c \overline c$ production below
the $D \overline D$ threshold, neglecting the color and spin of
the produced $c \overline c$ pair.  The leading order (LO) 
rapidity distributions of $J/\psi$'s produced in $dA$ collisions at impact 
parameter $b$ is
\begin{eqnarray} \frac{d\sigma}{dy d^2b d^2r} & = & 
2 F_{J/\psi} K_{\rm th} \int \,dz \, dz'
\int_{2m_c}^{2m_D} M dM \left\{ F_g^d(x_1,Q^2,\vec{r},z) 
F_g^A(x_2,Q^2,\vec{b} - \vec{r},z')
\frac{\sigma_{gg}(Q^2)}{M^2} \right. \nonumber \\ & & 
\mbox{} \left. \!\!\!\!\!\!\!\!\!\! \!\!\!\!\!\!\!\!\!\! \!\!\!\!\!\!\!\!\!\! 
+ \sum_{q=u,d,s} [F_q^d(x_1,Q^2,\vec{r},z) 
F_{\overline q}^A(x_2,Q^2,\vec{b} - \vec{r},z') +
F_{\overline q}^d(x_1,Q^2,\vec{r},z) F_q^A(x_2,Q^2,\vec{b} - \vec{r},z')] 
\frac{\sigma_{q \overline 
q}(Q^2)}{M^2} \right\}  \, \,  .
\label{sigmajpsi}
\end{eqnarray}
The LO partonic $c \overline c$ cross sections are defined in
Ref.~\cite{combridge}, $M^2 = x_1x_2s$ and $x_{1,2} =
(M/\sqrt{s_{NN}}) \exp(\pm y) \approx
(m_{J/\psi}/\sqrt{s_{NN}})\exp(\pm y)$ where $m_{J/\psi}$ is the
$J/\psi$ mass.  The fraction of $c \overline c$ pairs below the $D
\overline D$ threshold that become $J/\psi$'s, $F_{J/\psi}$, is fixed
at next-to-leading order (NLO) \cite{HPC}.  Both this fraction and the
theoretical $K$ factor, $K_{\rm th}$, drop out of the ratios.  We use
$m_c=1.2$ GeV and $Q = M$ \cite{HPC}.

We assume that the nuclear parton densities, $F_i^A$, are the product of
the nucleon density in the nucleus $\rho_A(s)$, the nucleon parton density
$f_i^N(x,Q^2)$, and a shadowing function $S^i_{{\rm P},{\rm S}}
(A,x,Q^2,\vec{r},z)$
where $\vec{r}$ and $z$ are the 
transverse and longitudinal location of the parton in position space with 
$s=\sqrt{r^2+z^2}$ and $s' = \sqrt{|\vec{b} - \vec{r}|^2 + z'^2}$.  
The first subscript, P, refers to the choice of shadowing parameterization,
while the second,  S, refers to the spatial dependence.
All available shadowing parameterizations ignore
effects in deuterium.  However, we take the proton and neutron
numbers of both nuclei into account.  Thus,
\begin{eqnarray}
F_i^d(x,Q^2,\vec{r},z) & = & \rho_d(s) f_i^N(x,Q^2) \label{fadeut} \\
F_j^A(x,Q^2,\vec{b} - \vec{r},z') & = & \rho_A(s') 
S^j_{{\rm P},{\rm S}}(A,x,Q^2,\vec{b} - \vec{r},z') 
f_j^N(x,Q^2) \, \, .  \label{fanuc} 
\end{eqnarray}
In the absence of nuclear modifications, $S^i_{{\rm P},{\rm
S}}(A,x,Q^2,\vec{r},z)\equiv 1$.  The nucleon densities of the heavy
nucleus are assumed to be Woods-Saxon distributions based on
measurements of the nuclear charge distributions with $R_{\rm Au} =
6.38$ fm and $R_{\rm Pb} = 6.62$ fm \cite{Vvv}.  The deuteron
wave function is predominantly $S$ wave with a small $D$ wave
admixture.  We use the Hulthen wave function\cite{hulthen} to calculate the
deuteron density distribution.  The central densities are normalized
so that $\int d^2r dz \rho_A(s) = A$. We employ the MRST LO parton
densities \cite{mrstlo} for the free nucleon.

We have chosen two recent parameterizations of the nuclear shadowing
effect which cover extremes of gluon shadowing at low $x$.  The Eskola
{\it et al.} parameterization, EKS98, is based on the GRV LO
\cite{GRV} parton densities.  Valence quark shadowing is identical for
$u$ and $d$ quarks.  Likewise, the shadowing of $\overline u$ and
$\overline d$ quarks are identical. Shadowing of the heavier flavor
sea, $\overline s$ and higher, is calculated separately.  The
shadowing ratios for each parton type are evolved to LO for $1.5 < Q <
100$ GeV and are valid for $x \geq 10^{-6}$ \cite{EKRS3,EKRparam}.
Interpolation in nuclear mass number allows results to be obtained for
any input $A$.  The parameterization by Frankfurt, Guzey and
Strikman, denoted FGS here, combines Gribov theory with hard
diffraction \cite{FGS}.  It is based on the CTEQ5M \cite{cteq5} parton
densities and evolves each parton species separately to NLO for $2 < Q
< 100$ GeV.  Although the given $x$ range is $10^{-5} < x < 0.95$, the
sea quark and gluon ratios are unity for $x > 0.2$.  The EKS98 valence
quark shadowing ratios are used as input since Gribov theory does not
predict valence shadowing.  The parameterization is available for four
different values of $A$: 16, 40, 110 and 206.  We therefore use the FGS
parameterization for $A =
206$ with both Au and Pb.

\begin{figure}[htb] 
\setlength{\epsfxsize=0.95\textwidth}
\setlength{\epsfysize=0.5\textheight}
\centerline{\epsffile{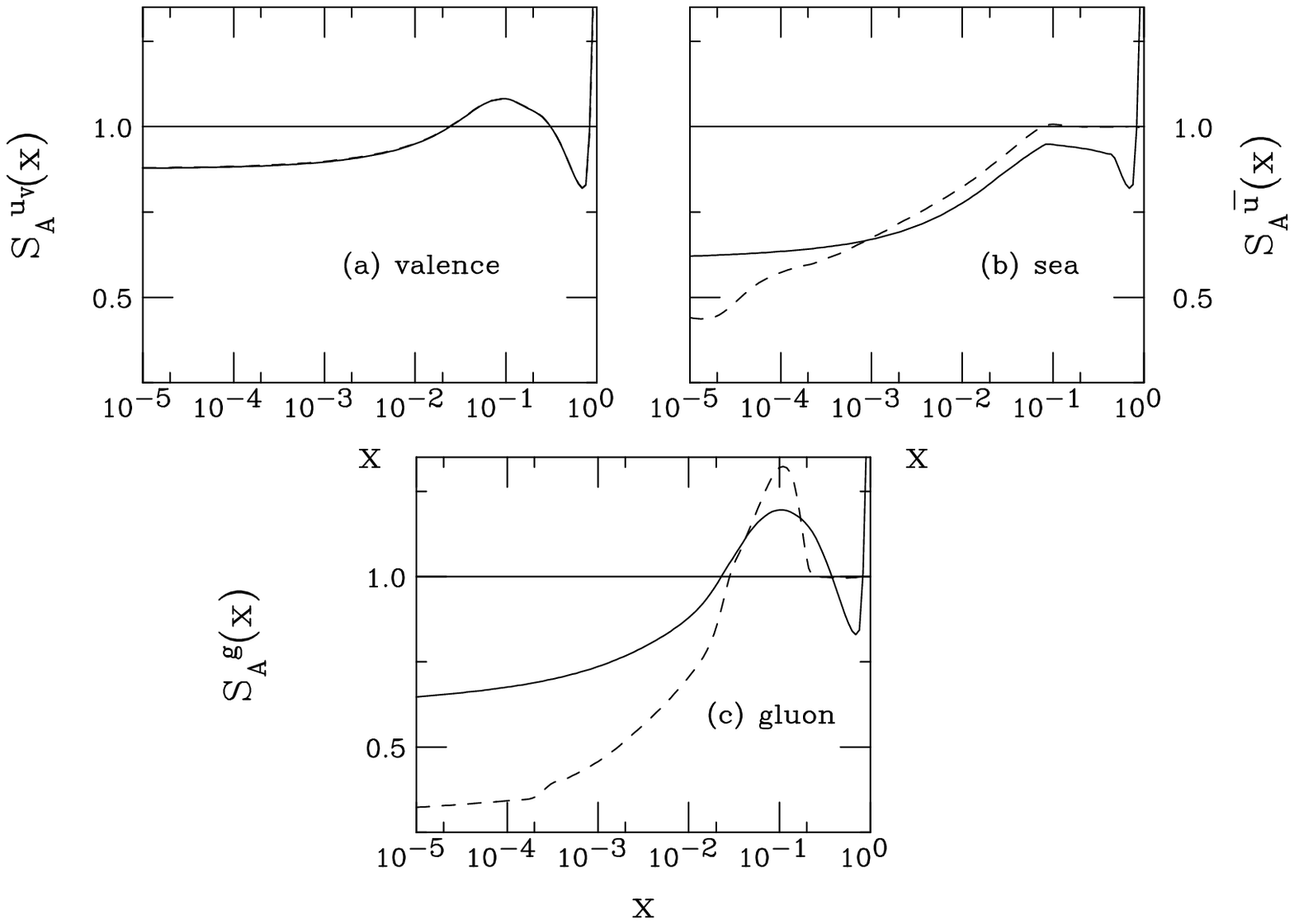}}
\caption[]{ The EKS98 and FGS shadowing parameterizations are compared at
the scale $\mu = 2m_c = 2.4$ GeV.  The solid curves are the EKS98
parameterization, the dashed, FGS.
}
\label{fshadow}
\end{figure}

Figure~\ref{fshadow} compares the homogeneous ratios, $S_{\rm EKS98}$
and $S_{\rm FGS}$ for $Q=2m_c$.  The FGS calculation predicts far more
shadowing at small $x$.  The difference is especially large for
gluons.  The FGS approach thus predicts larger antishadowing at $x
\sim 0.1$.  The valence and sea ratios are also shown, for
completeness.

We have calculated homogeneous shadowing effects on $J/\psi$
production at NLO.  However, the additional
integrals required to calculate inhomogeneous effects at NLO lead to
significant numerical difficulties.  Therefore we have compared the LO
and NLO calculations of homogeneous shadowing in $p$Au interactions at
$\sqrt{s_{NN}} = 200$ GeV using the EKS98 parameterization to verify that
the results are essentially identical. 

Some differences may arise from the use of the MRST HO \cite{mrstho}
parton densities in the NLO code while the LO calculation uses the
MRST LO \cite{mrstlo} parton densities.  In addition, at NLO, a new
production channel, $qg$ scattering, also contributes.  However, its
contribution is small.  The NLO and LO ratios agree well because
$J/\psi$ production is dominated by gluons for $x_F<0.6$.  Thus the $gg$
channel is dominant for $|y|<3$ at RHIC and $|y|<6.4$ at the LHC and
the LO calculation should reproduce the essential effects of
inhomogeneous shadowing.

Shadowing is not the only nuclear effect on $J/\psi$ production.
Broadening of the $p_T$ distributions, observed in fixed-target
measurements \cite{ptdat}, arises from multiple scattering of the
initial-state partons \cite{ptcalcs} and does not affect the total
cross section.  Thus, a $p_T$-integrated measurement should be
insensitive to this effect.  Nucleon absorption depends on the
$J/\psi$ production mechanism \cite{ramona}.  Absorption is generally
expected to increase at large negative rapidities where the $J/\psi$
may hadronize inside the target.  However, this regime is reached
outside the rapidity coverage of the present detector configurations.

We now turn to the spatial dependence of the shadowing.  Since the
data are limited, we use parameterizations of the spatial dependence.
We compare two parameterizations that we have used for inhomogeneous
shadowing in $AA$ collisions\cite{ekkv4,us,firstprl,spenpsi}.  The
first, $S_{{\rm P}, {\rm WS}}$, assumes that shadowing is proportional
to the local density, $\rho(s)$,
\begin{eqnarray}
S^i_{{\rm P},{\rm WS}}(A,x,Q^2,\vec{r},z) & = & 1 + N_{\rm WS}
[S^i_{\rm P}(A,x,Q^2) - 1] \frac{\rho_A(s)}{\rho_0} \label{wsparam} \, \, ,
\end{eqnarray}
where $\rho_0$ is the central density and $N_{\rm WS}$ is chosen so
that $(1/A) \int d^2r dz \rho_A(s) S^i_{{\rm P},{\rm WS}} = S^i_{\rm
P}$. When $r \gg R_A$, the nucleons behave as free particles while in
the center of the nucleus, the modifications are larger than the
average value $S^i_{\rm P}$.

If, on the other hand, shadowing stems from multiple interactions of
the incident parton\cite{ayala}, parton-parton interactions are spread
longitudinally over the coherence length, $l_c=1/2m_Nx$, where $m_N$
is the nucleon mass\cite{ina}.  For $x<0.016$, $l_c >R_A$ for any $A$.
The interaction is then delocalized over the entire trajectory so that
the incident parton interacts coherently with all the target partons
along its path length. At large $x$, $l_c\ll R_A$ and shadowing is
proportional to the local density, as above.  In practice, the
numerical differences between the two models are small and the
available data\cite{E745} is inadequate to tell the difference.
However, the both models predict significant differences with the
homogenous shadowing that is usually used.

Because of the difficulty of matching the shadowing at large and small
$x$ while conserving baryon number and momentum, we consider the small
$x$ and large $x$ limits separately.  This matching is necessary
because, for $y < -2$ at RHIC, $l_c < R_{\rm Au}$.  This regime is
measurable in the PHENIX muon spectrometers \cite{frawley}.
Equation~(\ref{wsparam}) corresponds to the large $x$ limit.  
When $l_c\gg R_A$, the spatial dependence may be parameterized as 
\be
S^i_{{\rm P},\rho}(A,x,Q^2,\vec{r},z) = 1 + N_\rho (S^i_{\rm P}(A,x,Q^2) - 1)
\frac{\int dz \rho_A(\vec{r},z)}{\int dz \rho_A(0,z)} \label{rhoparam} \, \, . 
\ee 
The integral
over $z$ includes the material traversed by the incident nucleon.  The
normalization requires $(1/A) \int d^2r dz \rho_A(s) S^i_{{\rm
P},\rho} = S^i_{\rm P}$. We find $N_\rho > N_{\rm WS}$.

\begin{figure}[htb] 
\setlength{\epsfxsize=0.95\textwidth}
\setlength{\epsfysize=0.5\textheight}
\centerline{\epsffile{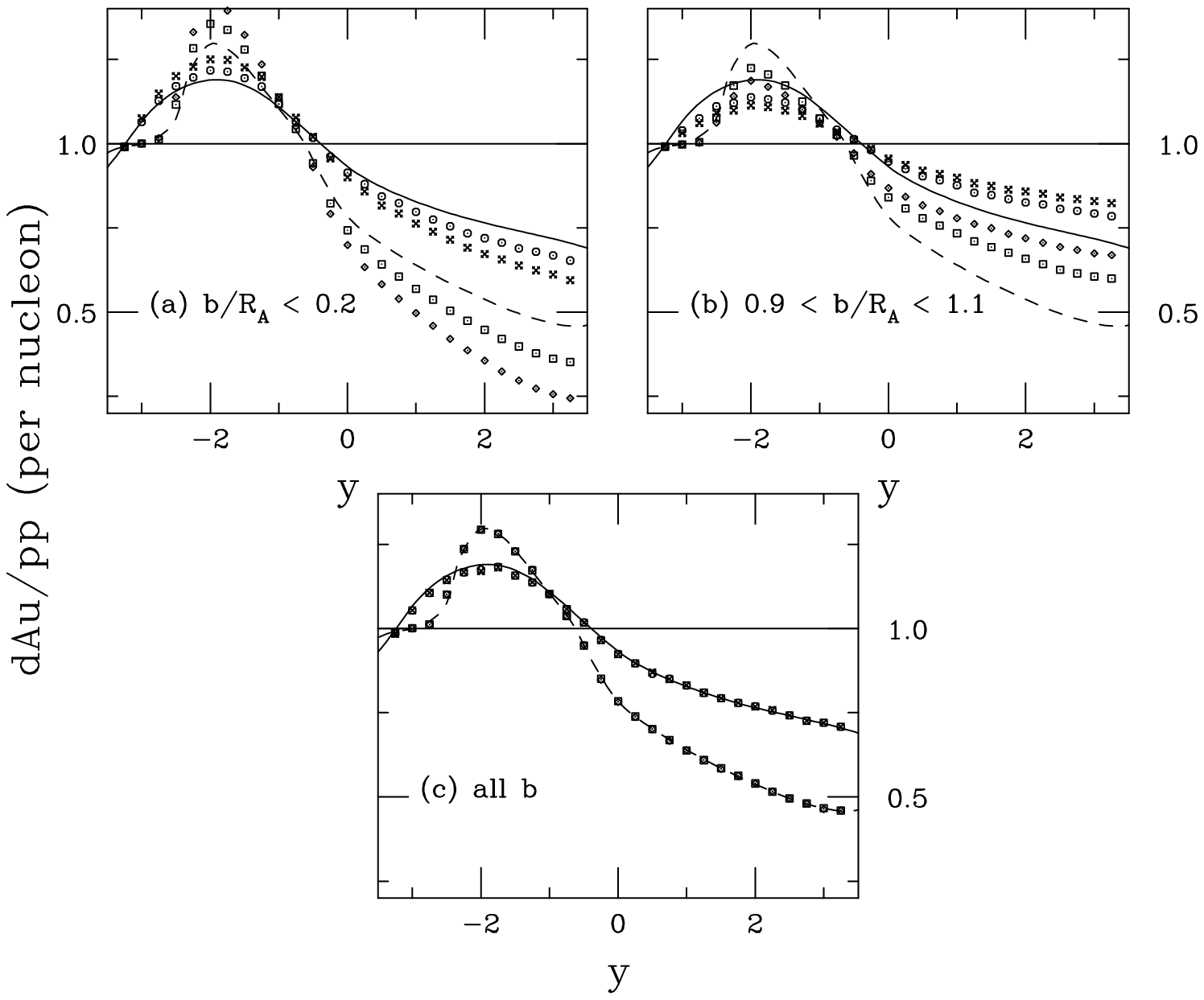}}
\caption[]{ The $J/\psi$ shadowing ratio at 200 GeV as a function of rapidity.
The results are shown for the EKS98 (solid line for homogeneous
shadowing, circles and x's for  $S^i_{{\rm EKS98},{\rm WS}}$ and  $S^i_{{\rm
EKS98},\rho}$ respectively) and FGS (dashed line for 
homogeneous shadowing, squares and diamonds for  $S^i_{{\rm FGS},{\rm WS}}$ 
and  $S^i_{{\rm FGS},\rho}$ respectively). The impact
parameter bins are (a) $b/R_A < 0.2$, (b) $0.9 < b/2R_A < 1.1$,
and (c) all $b$.}
\label{psiyrhic}
\end{figure}

\begin{figure}[htb] 
\setlength{\epsfxsize=0.95\textwidth}
\setlength{\epsfysize=0.5\textheight}
\centerline{\epsffile{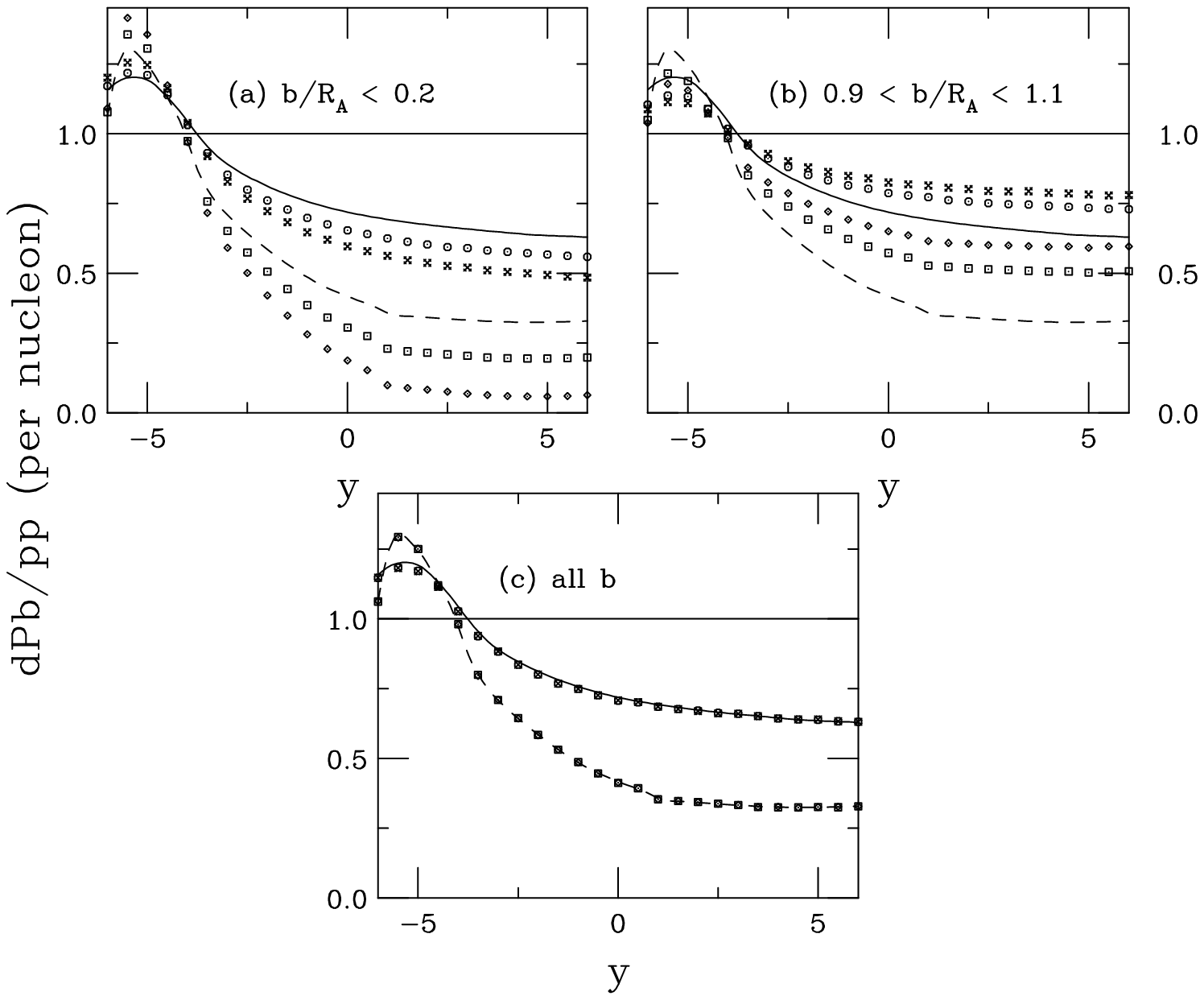}}
\caption[]{ The $J/\psi$ shadowing ratio at 6.2 TeV as a function of
rapidity.  The results are shown for the EKS98 (solid line for
homogeneous shadowing, circles and x's for $S^i_{{\rm EKS98},{\rm
WS}}$ and $S^i_{{\rm EKS98},\rho}$ respectively) and FGS (dashed line
for homogeneous shadowing, squares and diamonds for $S^i_{{\rm
FGS},{\rm WS}}$ and $S^i_{{\rm FGS},\rho}$ respectively). The impact
parameter bins are (a) $b/R_A < 0.2$, (b) $0.9 < b/2R_A < 1.1$, and
(c) all $b$.}
\label{psiylhc}
\end{figure}

Figures~\ref{psiyrhic} and \ref{psiylhc} show the ratios of $J/\psi$
production per nucleon in $dA$ interactions with and without shadowing
as a function of rapidity at RHIC ($\sqrt{s_{NN}} = 200$ GeV) and the
LHC ($\sqrt{s_{NN}} = 6.2$ TeV).  These ratios are essentially
equivalent to the $dA$ per nucleon to $pp$ ratio at the same energy
since isospin has a negligible effect on $J/\psi$ production.

We consider three cases: central collisions, $b<0.2R_A$, peripheral
collisions, $0.9R_A<b<1.1R_A$, and minimum-bias interactions, covering
all $b$.  Central collisions can be selected by choosing events with a
large number of `grey protons' which move with approximately the beam
rapidity.  Grey protons, slow in the target rest frame, are ejected
from the heavy nucleus in the collision.  The name originates from the
appearance of their tracks in nuclear emulsion.  Their momentum must
be larger than the Fermi momentum of the target.  The number of grey
protons is related to the number of individual $NN$ collisions, $\nu$,
which is closely related to the impact parameter.  Thus the number of
grey protons could be used to select quite central collisions.  This
technique was used in the $\nu A$ study of inhomogeneity \cite{E745}
as well as in lower energy $pA$ studies\cite{cole}.  In the multiple
interaction picture, shadowing should be directly proportional to
$\nu$.

At a heavy ion collider, grey protons can be detected in a forward
proton calorimeter (FPC).  Neutrons detected in a zero degree
calorimeter (ZDC), could also be used, although, because of the
transverse momentum, the acceptance would be limited.

The number of $NN$ collisions is larger in $dA$ than in
$pA$ interactions, reducing the uncertainty in the impact parameter
determination.  In a large acceptance detector, it
might also be possible to use total multiplicity to enhance the impact
parameter determination.

Peripheral collisions could be selected using two criteria.  Selecting
events with a small number of grey protons preferentially chooses a
small number of interactions.  It is also possible to select events
where only one of the nucleons in the deuteron interacts while the
other is unaffected.  Non-interacting neutrons can be detected in a
ZDC while non-interacting protons could be seen in a FPC.  In these
$NA$ collisions, the interacting deuterium nucleon is usually near the
surface of the heavy ion, so that $b>R_A - 2$ fm.  However, since the
deuteron has a significant density even at large distances, not all
single-nucleon interactions are very peripheral.

A complication for $dA$ collisions is that the difference between one
or two deuterium participants in the interaction is significant.  This
complication can be partly controlled by detecting the non-interacting
nucleon in forward calorimeters.  It can be further controlled by
studying the $J/\psi$ rapidity distribution.  The rapidity, $y=y_c$,
at which shadowing disappears, $S^i_{\rm P} \rightarrow 1$, could be
used as a calibration point.  If the deuterium comes from negative
rapidity, $J/\psi$ production at negative $y$ occurs via small $x$
partons in the deuteron.  The rapidity distribution ratios are
essentially mirror images of the shadowing parameterizations in
Fig.~\ref{fshadow}.  Since $S_{\rm P}^g \approx 1$ at $x \approx
0.025$, $y_c \approx -0.2$ at RHIC and -3.2 at the LHC.

The $J/\psi$ rate at $y=y_c$ could be used to normalize the cross
sections in different impact parameter classes, necessary if the
absolute efficiency of cuts on grey protons is difficult to assess.
Note that $y>y_c$ corresponds to low $x$ in the heavy nucleus while at
$y<y_c$, the antishadowing region is entered.  The point $y=y_c$ could
also be used to normalize the $dA$ data to $pp$ data at the same
energy, allowing a direct measurement of shadowing since higher order
corrections unrelated to shadowing cancel out in the ratio.  Although
$y_c$ may not be perfectly known, the exact value of $y_c$ is not
important since a slight error will only affect the absolute values of
the ratio.  Another advantage to this normalization is that, if, as
expected, the $J/\psi$ absorption is independent of rapidity in the
region of interest, any impact parameter dependence in the absorption
correction will drop out.

The homogeneous results are indicated by the solid (EKS98) and dashed
(FGS) lines.  The local density approximation of inhomogeneous
shadowing is shown by the circles and squares while the path length
approximation is shown by the x's and diamonds.  It is obvious that
$S^i_{{\rm P},\rho}$ has a stronger spatial dependence.

In central collisions, the inhomogeneous shadowing is stronger.  The
stronger the homogeneous shadowing, the larger the effect of the
spatial dependence.  In peripheral collisions, the inhomogenous
effects are somewhat weaker than the homogenous results but some
shadowing is still present.  Shadowing persists in part because the
density in a heavy nucleus is large and approximately constant except
close to the surface and partly because the deuteron wave function has
a non-zero amplitude at quite large distances.  Due to these tails, we
expect some shadowing effects even at very large impact parameters
where there are almost no interactions.  Figures~\ref{psiyrhic}(c) and
\ref{psiylhc}(c) show that the impact parameter integrated results are
identical to the homogenous results, as expected.

In $dA$ collisions, the rapidity-dependent inhomogeneity is somewhat
stronger than that predicted for $AA$ collisions\cite{ekkv4}.
Central $dA$ collisions are confined to the center of the target
nucleus, where the effect is stronger, while central $AA$ collisions
include both central and peripheral nucleon-nucleon interactions.

The predicted spatial dependence would be even stronger for proton
projectiles than for deuterons since the proton is effectively a point
particle in this approach.  In contrast, the deuteron is the same
size, $r\approx 2$ fm, as an intermediate mass nucleus.  However, it
would be considerably more difficult to determine the impact parameter
in $pA$ collisions since the number of grey protons is reduced and no
spectator nucleon would be left on the proton side.  In fact, light
ion projectiles may offer an attractive combination, with enough
interactions for an accurate impact parameter measurement, but little
projectile shadowing.

It should be possible to measure shadowing at midrapidity at RHIC,
particularly if the gluon shadowing is strong, as with the FGS
parameterization.  However, measurements at high $|y|$ could prove
more insightful, as seen in Fig.~\ref{psiyrhic}.  At RHIC,
measurements have been made in $pp$, $d$Au and Au+Au at the same
$\sqrt{s_{NN}}$, allowing direct comparison.  The STAR electromagnetic
calorimeter\cite{STAR} covers $-2<\eta<1$, encompassing part of the
antishadowing and shadowing regions.  The PHENIX muon
arms\cite{frawley} cover $1 \le |y| \le 2.4$ and could measure
the low and high $x$ regions while the PHENIX electron spectrometers,
$|y| \le 0.35$, cover the region near $y_c$.  It would then be
possible to compare the antishadowing region, $y \sim -2$ with the
shadowing region, $y\sim 2$.  Both detectors should be able to study
most of the $J/\psi$ $p_T$ spectrum.  A cut on minimum lepton $p_T$
might reduce the sensitivity to low $p_T$ $J/\psi$s.  A cut on
$J/\psi$ $p_T$ would increase the $x$ and $Q^2$ values probed at a
given $y$, allowing cross-checks on the shadowing
data.  The `cost' of this cut would be some sensitivity to $p_T$ broadening.

The higher energy of the LHC moves the antishadowing region out of the
reach of any planned lepton coverage. Only the low $x$ and transition
regions might be observable at low $p_T$, see Fig.~\ref{psiylhc}.
Both the central detectors of ALICE \cite{ALICE_central} ($|y| \leq
0.9$) and CMS \cite{CMS} ($|y| \leq 2.4$ for the central and forward
arms) should be able to clearly distinguish both the homogeneous
effect and the gross features of the inhomogeneity. The ALICE forward
muon spectrometer \cite{ALICE_muon}, $2.5 \leq y \leq 4$, could probe
the lowest and highest $x$ values possible at the LHC, if both $dA$
and $Ad$ collisions were studied. A $p_T$ cut on the $J/\psi$ would
increase the $x$ and $Q^2$ values, perhaps moving the antishadowing
region within reach.  The $\Upsilon$ has a very similar production
mechanism, but would probe larger $x$ and $Q^2$ values, and should thus
be sensitive to antishadowing at large $|y|$.

In conclusion, we have shown how nuclear shadowing can be measured
with $J/\psi$ production in $dA$ collisions.  The $J/\psi$ rapidity is
related to the momentum fraction of the target parton, allowing for
measurement of the $x$ dependence of shadowing.  By selecting events
with different numbers of grey protons and/or a surviving spectator
nucleon from the deuteron, it is possible to study the spatial
dependence of shadowing.  In all of the models considered, the spatial
dependence is large enough to be easily measured.  Although we have
focused on $J/\psi$, the approach used here should apply to other hard
probes such as $\Upsilon$, $W^{\pm}$, $Z^0$, Drell-Yan, open heavy
flavor and jet production.

We thank K.J. Eskola and V. Guzey for providing the shadowing routines
and for discussions.  This work was supported in part by the Division
of Nuclear Physics of the Office of High Energy and Nuclear Physics of
the U. S. Department of Energy under Contract Number
DE-AC03-76SF00098.

\end{document}